\documentclass[10pt,conference]{IEEEtran}
\IEEEoverridecommandlockouts
\usepackage{cite}
\usepackage{amsmath,amssymb,amsfonts}
\usepackage{bm}
\usepackage[noend]{algpseudocode}
\usepackage{algorithm}
\usepackage{graphicx}
\usepackage{textcomp}
\usepackage{xcolor}
\usepackage{mathtools}
\usepackage{fancyhdr}
\usepackage{multirow}
\usepackage{eso-pic}

\usepackage[nolist,nohyperlinks]{acronym}
\usepackage{enumitem}

\def\BibTeX{{\rm B\kern-.05em{\sc i\kern-.025em b}\kern-.08em
    T\kern-.1667em\lower.7ex\hbox{E}\kern-.125emX}}
\usepackage{nopageno}
\begin{document}

\pagestyle{empty}

\title{Reliability-Guided Trusted Repeater Node Selection in QKD-Enabled Metro Optical Networks}

\author{\IEEEauthorblockN{Arup Kumar Marik\IEEEauthorrefmark{1}, Basabdatta Palit\IEEEauthorrefmark{2}, and Sadananda Behera\IEEEauthorrefmark{1}} \\
\IEEEauthorblockA{\IEEEauthorrefmark{1} Dept. of Electronics and Communication Eng., National Institute of Technology Rourkela, Odisha, India\\
\IEEEauthorrefmark{2} Dept. of Information Technology, Indian Institute of Engineering Science and Technology, Shibpur, Howrah, India}}

\maketitle

\thispagestyle{empty}
\begin{abstract}
Quantum Key Distribution (QKD) in optical networks can provide information-theoretic security but is limited by its operability range, thereby requiring repeater nodes for extended coverage. Existing works typically assume that all repeater nodes and their associated key management systems are fully trusted, overlooking the risks posed by software exploits and insider threats. To address this, in this work, we propose a Bayesian fusion-based model to quantify the trustworthiness of each node. We have formulated a reliability-guided Trusted Repeater Node (TRN) selection framework for metro optical networks, where each node is assigned a trust score. We have subsequently mapped this trust score to link weights, which in turn has been used for reliable path computation for TRN selection, using the Dijkstra algorithm. We have also ranked the nodes using a composite score combining eigenvector and betweenness centrality, to capture their topographical relevance in the network. Simulation results on a reference topology demonstrate that the proposed method achieves approximately 8.93\% higher path coverage than traditional centrality-based approaches like degree centrality, using the same number (around ten) of TRNs, 
thereby supporting reliable and resilient TRN selection for QKD-enabled optical networks.

\end{abstract}

\acresetall
\section{Introduction}
Quantum computing technologies have gained significant attention in recent years owing to their potential to compromise the security of conventional public-key cryptographic algorithms such as RSA~\cite{Rivest1978}, whose security relies on the computational hardness of the integer factorization problem. In particular, recent advances in quantum hardware, error correction, and algorithmic optimization have accelerated the practical realization of Shor's algorithm~\cite{Shor_1997}, thereby raising concerns regarding the long-term resilience of classical cryptographic methods. To address these challenges, Quantum Key Distribution (QKD) has emerged as an information-theoretically secure approach for cryptographic key distribution~\cite{BENNETT2014}. However, the practical deployment of QKD over optical fiber networks is constrained by transmission distance limitations, typically extending to only a few hundred kilometers~\cite{cao2026experimental}, primarily due to signal attenuation and detector inefficiencies.

To facilitate long-distance secure communication in the absence of quantum amplifiers, intermediate relay or repeater nodes are employed. These nodes receive, decrypt, and retransmit the secret keys to successive links, enabling the distribution of quantum keys over longer distances. However, as these relays process the key using electronic systems, they present potential security risks. Hence, ensuring that such intermediary nodes remain secure and trustworthy is critical for preserving the overall confidentiality of the QKD network.

% Early works have explored TRN deployment using heuristic strategies such as Minimum Spanning Tree (MST) and shortest-path-based routing \cite{Gunkel2019}, aiming to minimize link distances and centralize key distribution. However, these approaches failed to ensure optimal key rates due to link congestion and lacked resilience, making them unsuitable for robust metro-scale deployment. To reduce TRN count while maintaining secure connectivity, \cite{Patri2023} proposed a heuristic method which used span aggregation and multi-device configurations per node. Despite its innovative integration of multiple fiber pairs, the approach assumes ideal device availability and does not consider node-level vulnerabilities or reliability variations. Most existing optimization-based TRN planning methods assume that relay node locations are predetermined and fully trusted \cite{Pederzolli2020}. This assumption oversimplifies real-world scenarios, where node reliability may vary due to administrative control, physical access, or cyber threats. These methods also face scalability challenges due to the exponential growth in solution space, leading to high computational overhead.
Several research efforts have focused on optimizing the placement of \acp{TRN} in QKD-enabled optical networks. Early studies primarily employed graph-theoretic heuristics such as the Minimum Spanning Tree (MST) and Single-Source Shortest Path Tree  algorithms for \ac{TRN} selection~\cite{Gunkel2019}. Although computationally simple, these methods often suffer from link congestion, limited resilience, and reduced secure key rates (SKR). A cost-efficient quantum network design using Steiner-tree-based heuristic algorithm was presented in~\cite{Ilora2023}, which minimizes \ac{TRN} and dark-fiber deployment while outperforming MST-based approaches in cost efficiency. However, it overlooks node reliability and physical security aspects that are essential for trustworthy network operation. A mixed-integer linear programming (MILP) approach was introduced in~\cite{Pederzolli2020} to minimize fiber utilization; however, it assumed predetermined and fully trusted \ac{TRN} locations, limiting its practical applicability. Authors in~\cite{Patri2023} proposed a span aggregation algorithm  to reduce the number of \acp{TRN} in optical transport networks, achieving notable fiber cost savings across different topologies. However, this design introduces a trade-off between cost efficiency and SKR, and it does not account for node-level vulnerabilities arising from classical processing in \acp{TRN}. In~\cite{Rabbie2022}, authors formulated the repeater allocation problem as an integer linear program (ILP) that minimizes the number of quantum repeaters while maintaining rate, fidelity, and robustness constraints. Although scalable to realistic network sizes, the ILP approach is computationally demanding and assumes limited end-node activity. A two-step ILP method proposed in~\cite{Romtham2024} reduces computational complexity but continues to rely on static end-node assumptions. More recently, a traffic-aware \ac{TRN} placement strategy~\cite{Dibaj2025} based on the Hot-Link algorithm was proposed to adaptively allocate trusted nodes according to SKR demand and asymmetric traffic, improving utilization compared to static schemes.

Most of these works have primarily focused on TRN placement in backbone optical networks and, in the existing literature~\cite{Gunkel2019,Ilora2023,Patri2023,Pederzolli2020}, assume that these relay nodes are entirely trustworthy~\cite{Kong2024}, an assumption that is often impractical in realistic environments. The trustworthiness or reliability of nodes depends on their physical safeguards, firewall configurations, administrative oversight, and exposure to network-based threats. Therefore, it is crucial to incorporate node reliability into the design and selection of \ac{TRN}s to guarantee the end-to-end resilience and security of the QKD infrastructure.

To address this gap, the authors in~\cite{marik2025} proposed a reliability-aware TRN selection framework for metropolitan area networks by incorporating node reliability into path-weight computation. However, the node reliability values were assigned randomly (0.5–1.0) rather than derived from a reliability model. In addition, the framework relied on multiple tunable parameters for path weighting and TRN ranking, requiring retuning for different network topologies and reliability distributions. 

In contrast, the present work develops a systematic, simulation-driven reliability modelling framework based on \textbf{\textit{Bayesian fusion method}}~\cite{sander2013}, providing a probabilistic approach for estimating node reliability. Rather than assigning reliability values arbitrarily, the proposed model quantifies node trustworthiness by aggregating multiple operational and security indicators into a unified reliability score. The estimated reliability values are then directly integrated into the TRN selection process, eliminating the need for manually tuned weighting parameters. Consequently, the proposed framework enables automated, topology-agnostic TRN selection for QKD-enabled optical networks and can be readily adapted to diverse network configurations without parameter retuning.

% \textcolor{blue}{To address this gap, the authors in~\cite{marik2025} introduced a reliability-aware \ac{TRN} selection framework for metropolitan area networks that incorporated node trustworthiness into path-weight computation.
% However, the authors do not focus on reliability modelling; the node reliability values were assigned randomly, and the framework relied on several tunable parameters to balance distance and reliability in the path weights and to combine centrality scores for TRN ranking. These parameters were network-dependent and required retuning whenever the topology or reliability distribution changed, thereby limiting the model’s interpretability and scalability. In contrast, the present work develops a systematic, simulation-driven reliability model using a \textbf{\textit{Bayesian fusion}} approach~\cite{sander2013}, which provides a probabilistic basis for evaluating node trustworthiness. As a result, the proposed model can quantify the node reliability by aggregating multiple operational and security indicators, enabling a fully automated \ac{TRN} selection process for QKD-enabled optical networks, which is topology agnostic and, hence, adaptable over a wide variation of network configurations.}

The key contributions of this work can, therefore, be summarized as follows:
\begin{enumerate}
    \item We have adopted a Bayesian fusion-based reliability modeling approach to estimate the reliability of nodes in QKD-enabled optical networks by combining multiple operational and security indicators into a unified probabilistic reliability score.
    \item We have developed a reliability-aware \ac{TRN} selection framework that incorporates the estimated node reliability into path-weight computation and combines it with centrality-based ranking to identify secure and reliable \acp{TRN}.
    
    % Bayesian fusion is used to estimate reliability as a function of multiple operational and security indicators, which are then embedded into a modified Dijkstra algorithm to identify secure and reliable transmission routes in QKD-enabled optical networks. 
  
    % \item \textcolor{red}{We introduce Reliability Contribution (RC) as a reliability-aware evaluation metric and jointly use it with Path Coverage (PC) to assess the structural importance and operational trustworthiness of candidate TRNs.}
    % \item we validate the proposed framework on a 28-node metropolitan optical network and demonstrate its superiority over conventional centrality-based \ac{TRN} selection approaches.
    
     % \item \textcolor{blue}{We have introduced a new metric, Reliability Contribution (RC), and Path Coverage (PC) metric to jointly evaluate node reliability and structural importance. This integration enhances \ac{TRN} selection compared with traditional centrality-based methods, improving both key-distribution efficiency and network robustness.}
\end{enumerate}

We have validated the proposed framework on a 28-node metro optical network and demonstrate its effectiveness over conventional centrality-based approaches through comprehensive simulation studies.
Simulation results demonstrate the effectiveness of the proposed framework through comprehensive performance evaluation using reliability- and connectivity-based metrics. By jointly considering node reliability and structural centrality, the proposed method achieves approximately 8.93\% higher path coverage than degree centrality while using the same number (around ten) of \acp{TRN}.

% We have  validated the proposed framework on a 28-node metro optical network and demonstrate its effectiveness through comprehensive performance evaluation using reliability- and connectivity-based metrics. Simulation results show that jointly considering node reliability and structural centrality enables more effective \ac{TRN} selection, achieving approximately 8.93\% higher path coverage than degree centrality while using the same number (around ten) of \acp{TRN}.

\section{Reliability Modeling Using Bayesian Fusion Method} 
This section introduces the proposed Bayesian fusion-based reliability model designed to quantify the trustworthiness of nodes in \ac{QKD}-enabled optical networks. We first describe the overall system model, followed by the proposed Bayesian fusion framework, and finally discuss the implementation of the framework. 

\subsection{System Model}
The QKD network is represented as an undirected graph $G(V, E)$, where $V$ denotes the set of nodes and $E$ the set of optical fiber links. Each link $(u,v)\in E$ connects nodes $u$ and $v$ with a physical distance $d_{uv}$. The objective is to assign a reliability score $R_v$ to each node $v \in V$, which will later be used for reliability-aware \ac{TRN} selection to rank potential nodes for key relay operation and support secure path computation.

%\subsection{Motivation}
In contrast to previous works, we have considered the node reliability to depend on multiple operational and security-related parameters, such as:
\begin{itemize}
    \item Patch Latency (PL): Average delay in applying critical updates; lower latency implies better maintenance.
    \item Uptime (UP): Fraction of time the node remains operational; higher uptime denotes greater stability.
    \item Hardware Security (HS): Presence of Trusted Platform Module, secure boot, or hardware encryption.
    \item Physical Access (PA): Degree of physical protection, surveillance, and restricted access.
    \item Firewall Score (FS): Strength of network-level defenses, including intrusion detection/prevention systems  and access policies.
\end{itemize}
These indicators collectively form the evidence set $\{X^i_v\}$ for each node $v$.  We have assigned trust scores using a Bayesian fusion-based reliability model that systematically integrates these diverse reliability indicators %such as patch latency (PL), uptime (UP), hardware security (HS), physical access control (PA), and firewall score (FS) 
into a unified probabilistic estimate for each node~\cite{sander2013}.
% We have used Bayesian fusion as it naturally combines multiple heterogeneous and uncertain reliability indicators into a single probabilistic trust estimate.

\subsection{Bayesian Framework}
% Bayesian inferencing is conventionally used to integrate uncertain or incomplete information from multiple independent indicators into a unified probabilistic estimate~\cite{sander2013}. 
In order to implement Bayesian inferencing, we have modeled each node $v$ as a binary random variable $H_v \in \{H_1, H_0\}$, where $H_1$ denotes that the node is secure and $H_0$ indicates that it is compromised. The prior probabilities $P(H_1)$ and $P(H_0)=1-P(H_1)$ represent the initial belief about a node’s trustworthiness. Here, $P(H_1)$ is a tunable parameter.

For each node, the set of measurable reliability indicators $\{X^i_v\}$,  normalized to the range $[0,1]$, serves as evidence, i.e., 
\begin{equation}
\mathcal{X}_v = \{X^1_{v}, X^2_v, X^3_v, X^4_v, X^5_v\} = \{\text{PL}, \text{UP}, \text{HS}, \text{PA}, \text{FS}\},
\end{equation}
These reliability indicators, $X^i_v \in [0,1]$, will vary randomly due to some external factors. So, we have modeled $X^i_v$'s to take random values from a beta distribution, which is suitable for representing bounded uncertainty~\cite{fang2016btres}. Two beta distributions are defined per indicator, one under the secure hypothesis and another under the compromised hypothesis:
\begin{equation}
P(X^i_v|H_1) \sim \text{Beta}(\alpha_{i1}, \beta_{i1}), \quad
P(X^i_v|H_0) \sim \text{Beta}(\alpha_{i0}, \beta_{i0}).
\end{equation}

The beta distribution parameters were selected to represent realistic secure and compromised node behaviors while maintaining a partial overlap between the two hypotheses. This overlap allows the Bayesian inference process to distinguish trustworthy and vulnerable nodes without producing trivial classifications, thereby yielding meaningful posterior reliability estimates.
% The parameters $(\alpha_{i1}, \beta_{i1})$ and $(\alpha_{i0}, \beta_{i0})$ are extensively tuned to generate posterior reliability values within a realistic range ($0.5 \leq R_v \leq 1$) while ensuring a partial overlap between the two distributions. 
A minimum reliability value  of $0.5$ implies that the node behaviour is highly uncertain making it only  marginally reliable, while the maximum value $1$ implies a fully secure node.
% \subsection{Reliability Indicators}
% Following are the indicators for reliability estimation:
% Since their influence on node trustworthiness may vary, a probabilistic fusion is required to combine them into a single reliability estimate.
In the next section, we explain how we have used the Bayesian inference model to compute the posterior reliability score $R_v$ based on the observed reliability indicators.

\subsection{Posterior Reliability Estimation}
Given the observed set of indicators $\mathcal{X}_v^i$ for a node $v$, the posterior probability representing the node's reliability score is computed using Bayes’ theorem as:
% \begin{equation}
% \begin{aligned}
% R_v &= P(H_1|X^1_v, X^2_v, \ldots, X^n_v) \\[3pt]
% &= \frac{P(H_1)\prod_{i=1}^{n} P(X^i_v|H_1)}
% {P(H_1)\prod_{i=1}^{n} P(X^i_v|H_1) + P(H_0)\prod_{i=1}^{n} P(X^i_v|H_0)}.
% \end{aligned}
% \end{equation}
\begin{equation}
\begin{aligned}
R_v &= P(H_1|X^1_v, X^2_v, \ldots, X^n_v) \\[3pt]
&= \frac{P(H_1)\prod\limits_{i=1}^{n} P(X^i_v|H_1)}
{P(H_1)\prod\limits_{i=1}^{n} P(X^i_v|H_1) + P(H_0)\prod\limits_{i=1}^{n} P(X^i_v|H_0)}.
\end{aligned}
\end{equation}
Here, the numerator represents the joint likelihood that node $v$ is secure given the observed evidence. The resulting reliability score $R_v$ lies within the interval $[0.5, 1]$, where higher values indicate greater trustworthiness. %A lower bound of $R_v = 0.5$ serves as the minimum acceptable threshold for node participation in QKD key distribution or \ac{TRN} placement.

\subsection{Implementation and Computation}
The proposed Bayesian reliability framework was implemented in Python using the \texttt{NumPy} and \texttt{SciPy} libraries, and the node reliability was computed sequentially for each node as follows:
\begin{enumerate}
    \item For simulation purposes, all reliability indicators $X^i_v\in \mathcal{X}$ are assumed to be independent and identically distributed (i.i.d.) random variables generated from a Beta(5,3) distribution
    
    \item Likelihoods $P(X^i_v|H_1)$ and $P(X^i_v|H_0)$, based on the previously computed $X^i_v$ values, are obtained using the parameterized beta distributions: $\text{Beta}(4,3)$ for the secure hypothesis ($H_1$) and $\text{Beta}(6,7)$ for the compromised hypothesis ($H_0$), capturing both reliability and uncertainty in the node behaviour. As mentioned earlier, the shape parameters of these beta distributions have been extensively tuned to ensure that the reliability scores lie between 0.5 and 1.
    \item The joint likelihood is obtained by assuming conditional independence among the indicators.
   \item  The prior probability of a node being secured is uniformly set as $P(H_1)=0.7$ for all nodes. A prior probability of 0.7 is used because metro optical networks are generally expected to operate securely, while still accounting for the possibility of occasional node failures or security compromises. Because the prior is identical for all nodes, the relative TRN ranking is governed by the node-specific evidence rather than the choice of prior.
    \item Finally, Bayes’ theorem is applied to compute the posterior reliability $R_v$ for each node.
\end{enumerate}
% \begin{itemize}
%     \item Simulated evidence $X^i_v$ values for all reliability indicators are generated using a moderate $\text{Beta}(5,3)$ distribution to represent realistic operational conditions. Each indicator is then evaluated under two parameterized beta distributions: $\text{Beta}(4,3)$ for the secure hypothesis ($H_1$) and $\text{Beta}(6,7)$ for the compromised hypothesis ($H_0$), capturing both reliability and uncertainty in node behaviour.
%     \item Likelihoods $P(X^i_v|H_1)$ and $P(X^i_v|H_0)$ are computed using these parameterized distributions.
%     \item The joint likelihood is obtained by assuming conditional independence among the indicators.
%     \item Finally, Bayes’ theorem is applied to compute the posterior reliability $R_v$ for each node.
% \end{itemize}
% \textcolor{red}{To ensure that the computed reliability values are practically meaningful, the Beta distribution parameters were carefully tuned such that the posterior reliability scores $R_v$ remain within the interval $[0.5, 1]$. This threshold reflects a minimum acceptable security level of $50\%$, representing the lower bound of operational trustworthiness considered tolerable within the network.
% We apply this framework to compute $R_v$ for each node in a $28$-node optical network topology (Fig.~\ref{fig:network_topology}). The prior probability is uniformly set as $P(H_1)=0.7$ for all nodes. The resulting reliability values $R_v$ vary smoothly within the defined range, reflecting heterogeneous operational and security conditions across the network.}
The computed posterior reliability scores $R_v$ thus provide a quantitative measure of node trustworthiness and are subsequently integrated into the path-weight computation for reliability-aware \ac{TRN} placement.% To further capture the structural significance of each node, a composite trust score is constructed by combining betweenness and eigenvector centralities on the reliability-weighted graph, enabling the identification of top-$K$ nodes as optimal \acp{TRN}. Algorithm~\ref{alg:trn_placement} presents the pseudo-code of the proposed methodology.}

\section{Reliability-Aware Path Weight Modification}
In this section, we first explain how we have incorporated the reliability scores obtained from the proposed Bayesian fusion method into link weights. We then explain the proposed algorithm for ranking the TRNs using betweenness centrality and eigenvector centrality, based on modified Dijkstra's method.
\subsection{Weight Modification with Reliability (Line 2, Algorithm 1)}
Each link $(u, v)$ is assigned a modified weight $w'_{uv}$ based on the physical distance $d_{uv}$ and the combined reliability of its endpoints $(R_u \cdot R_v)$. The link weight is calculated as:
\begin{equation}
\label{weight}
w_{uv}' = \frac{d_{uv}}{R_u \cdot R_v}.
\end{equation}
Taking the inverse of $(R_u \cdot R_v)$  penalizes links between less reliable nodes. The modified weights are used to generate a new graph $G'(V, E)$. We apply Dijkstra's algorithm on this new graph to find the shortest path between all the source and destination pairs. Next, to capture the structural significance of each node based on the connections provided by Dijkstra's algorithm, a composite trust score is constructed by combining betweenness and eigenvector centralities on the reliability-weighted graph, enabling the identification of top-$K$ nodes as optimal \acp{TRN}. Algorithm~\ref{alg:trn_placement} presents the pseudo-code of the proposed methodology.  

\subsection{Node Ranking using Centrality Measures}
For each node $v \in G'(V, E)$ we compute the following centrality scores.
\begin{enumerate}
    \item \textbf{Betweenness Centrality (BC)}  quantifies the fraction of shortest paths in $G'(V, E)$ (computed using modified weights) that pass through node $v$, such that,
    \begin{equation}
    \label{BC}
    BC_v = \sum_{s \ne v \ne t} \frac{\sigma_{st}(v)}{\sigma_{st}},
    \end{equation}
    where $\sigma_{st}$ is the total number of shortest paths from source $s$ to destination $t$, and $\sigma_{st}(v)$ is the number of such paths that pass through $v$ (excluding endpoints).
    
    \item \textbf{Eigenvector Centrality (EC)} measures the influence of node $v$ based on the scores of its neighbors.  
    %For the modified reliability-weighted graph $G'(V, E)$ with $|V|$ vertices, 
    Let $A=(a_{v,z})$ denote the adjacency matrix, where each element $a_{v,z}$ corresponds to the inverse of the \textit{weight} of the link between vertices $v$ and $z$ in $G'$. Specifically, $a_{v,z} = 1/ w_{vz}'$ if vertex $v$ is connected to vertex $z$, and $a_{v,z} = 0$ otherwise. 
    The relative eigenvector centrality score, $EC_v$, of vertex $v$ is then defined as:
    \begin{equation}
    \label{EC}
    EC_v = \frac{1}{\lambda} \sum_{z \in \mathcal{N}(v)} a_{v,z} EC_z
    \end{equation}
    where $\mathcal{N}(v)$ is the set of neighbors of $v$, and $\lambda$ is a constant. A high $EC_v$ implies connections to other influential nodes.
\end{enumerate}

A new composite score, which determines the importance of the node based on its topological location and its influence on neighbouring nodes, is defined as follows:
\begin{equation}\label{eq:trust_score}
\text{TS}_v = BC_v\cdot EC_v
\end{equation}
 Nodes with the highest $\text{TS}_v$ are considered ideal candidates for the \ac{TRN}. This formulation ensures that a node receives a high score only when it is simultaneously central to shortest-path routing and well connected to other influential nodes.

\subsection{Ranking-Based \ac{TRN} Selection Algorithm}

\vspace{1mm}
\begin{algorithm}[t]
\caption{Ranking of Potential \acp{TRN} in a Network}
\label{alg:trn_placement}
\begin{algorithmic}[1]
\item[]  \textbf{Input:} Network topology \( G(V, E) \), node reliabilities \( R_v \), link distances \( d_{uv} \)
\item[]  \textbf{Output:} Ordered list of nodes for \ac{TRN}
\ForAll{edge \( (u, v) \in E \)}
    \State Compute modified weight: 
    
    \( w_{uv}' = \frac{d_{uv}} {(R_u \cdot R_v)} \)
\EndFor
\State Construct modified graph \( G'(V, E) \).

% \State Run Dijkstra's shortest path algorithm on the modified graph $G'$ using the modified link weights.

\ForAll{node \( v \in V \)}
    \State Compute \( BC_v \) on \( G' \).
    \State Compute  \( EC_v \) on \( G' \).
    \State Compute  \( \text{TS}_v =  BC_v\cdot EC_v \).
\EndFor

\State Rank all nodes in descending order of \( \text{TS}_v \).
\State \Return Top \( K \) nodes as potential \acp{TRN}.
\end{algorithmic}
\end{algorithm}

We then rank the nodes based on the reliability-aware composite total score ($TS_{v}$) evaluated in the previous step. 
%This method integrates node reliability with structural centrality to prioritize \ac{TRN} placement in QKD networks.
By weighting shortest-path computations with reliability and combining BC and EC, we identify nodes that are both structurally critical and trustworthy, supporting secure and efficient key distribution across metro optical networks.

The computational complexity of Algorithm~\ref{alg:trn_placement} is dominated by the betweenness centrality computation in weighted graphs, resulting in an overall complexity of $O(|V||E| + |V|^2 \log |V|)$. Hence, the algorithm is computationally feasible for metro-scale optical networks.

% however, scalability to very large networks may require approximation techniques or parallelization strategies.

\begin{figure}[t]
    \centering
    \includegraphics[width=0.5\textwidth]{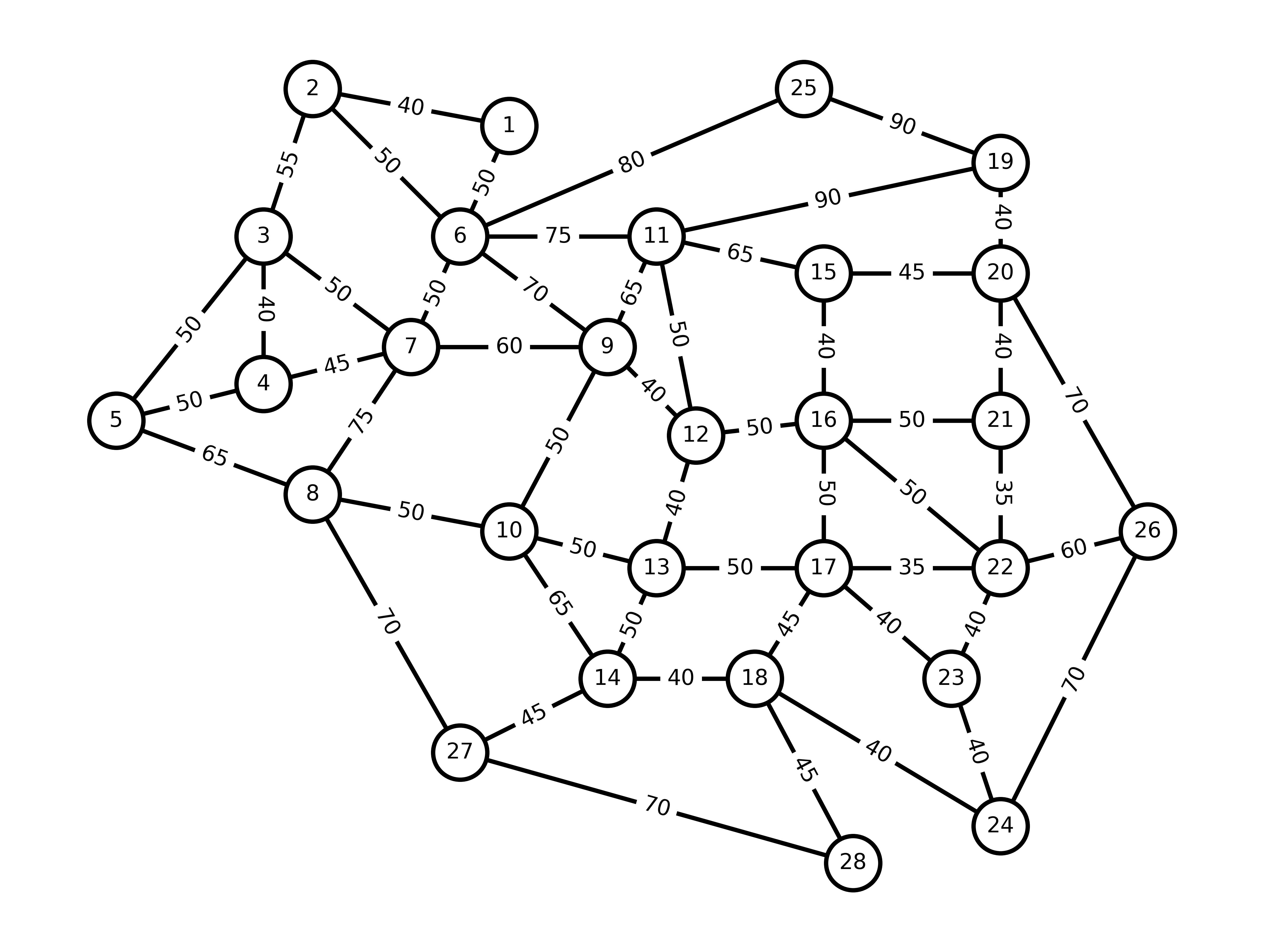}
\vspace{-.3in}\caption{Topology of a reference metro optical network, where the link distances    are  in kilometers.}
    \label{fig:network_topology}
\end{figure}
\section{Results and Discussions}
In this section, we have evaluated the proposed reliability-aware \ac{TRN} ranking framework on a reference metropolitan optical network topology~\cite{Yan2018}, consisting of $28$ nodes interconnected by $52$ bidirectional fiber links.

For each node $v$, the total score $TS_v$, which captures a node's structural importance and influence,  is computed as \eqref{eq:trust_score}. We have carried out the total score computation over $1000$ different random reliability values generated using the Bayesian fusion method described in Section~II. For every instance, the corresponding link weights are updated according to the reliability-adjusted formulation (see line~2 of Algorithm~1), and both PC and RC metrics are computed for all nodes. 

The final average values of PC, RC, and the total score $TS_v$ are obtained by averaging across all instances, following the procedure outlined in Algorithm~\ref{alg:trn_placement} (see line~7). Nodes are subsequently ranked in descending order of their averaged $TS_v$ values to identify the most suitable candidates for \ac{TRN} selection.

% All computations and graph-theoretic analyses were implemented using the \texttt{NetworkX} library in Python.

% For reliability computation of each node, we generated simulated evidence $X^i_v$ values for all reliability indicators using a moderate $\text{beta}(4,3)$ distribution to represent realistic operational conditions. We then computed the likelihoods $P(X^i_v|H_1)$ and $P(X^i_v|H_0)$ using the parameterized beta distributions for secure and compromised cases. The joint likelihood was evaluated by assuming conditional independence among the indicators. Then, Bayes’ theorem was applied to compute the posterior reliability $R_v$ for each node.

% The objective is to identify optimal node locations for \ac{TRN} deployment by jointly considering their structural centrality and reliability contribution.

\subsection{Node Ranking for \ac{TRN} Selection}
% For each node $v$, the total score $TS_v$ defined in~\eqref{eq:trust_score} is computed using its eigenvector and betweenness centralities, and nodes are subsequently ranked in descending order of $TS_v$.
Table~\ref{tab:top10_nodes} lists the top ten nodes identified as the most suitable candidates for \ac{TRN} placement. The results indicate that nodes exhibiting both high connectivity and strong reliability tend to achieve higher total scores. Among them, Node~9 attains the maximum $TS_v$ value of $0.062968$, primarily due to its strategic network position (high $BC_v$) and strong influence from well-connected neighboring nodes (high $EC_v$). It maintains direct connections with Nodes~6, 7, 10, 11, and~12, all of which also appear within the top ten rankings, thereby enhancing its topological prominence (see Fig.~1 and Table~I). 
Although Table~\ref{tab:top10_nodes} reports the top ten ranked nodes, the number of \acp{TRN} selected for key relay operation can be adapted based on specific network coverage or redundancy requirements. It is worth noting that only the selected \acp{TRN} are used for key relay operations along a path, while the remaining nodes serve as Optical Bypass (OB) nodes~\cite{dong2020auxiliary}.
% \vspace{-0.1in}
\begin{table}[ht]
\centering
\caption{Set of Top 10 \ac{TRN} Nodes Ordered According to Their Total Score.}
\begin{tabular}{|c|c|c|c|}
\hline
\textbf{Ranked Node} & \textbf{Total Score} & \textbf{Ranked Node} & \textbf{Total Score} \\
\hline
9  & 0.062968 & 10 & 0.017679 \\
\hline
6  & 0.053521 & 16 & 0.015220\\
\hline
7 & 0.031868 & 13 & 0.015009 \\
\hline
12 & 0.029856 & 17 & 0.014975\\
\hline
11 & 0.027565   & 8 & 0.013445 \\
\hline
\end{tabular}
\label{tab:top10_nodes}
\end{table}

% \begin{table}[ht]
% \centering
% \caption{Set of Top 10 \ac{TRN} Nodes Ordered According to Their Total Score.}
% \begin{tabular}{|c|c|}
% \hline
% \textbf{Ranked Node} & \textbf{Total Score} \\
% \hline
% 6  & 0.089200\\
% \hline
% 11  & 0.061496 \\
% \hline
% 9 & 0.044173  \\
% \hline
% 8 & 0.030312 \\
% \hline
% 10 & 0.026539  \\
% \hline
% 2 & 0.015731 \\
% \hline
% 3 & 0.009676 \\
% \hline
% 14 & 0.009611\\
% \hline
% 15 & 0.008805\\
% \hline
% 19 & 0.008266\\
% \hline
% \end{tabular}
% \label{tab:top10_nodes}
% \end{table}
% \vspace{-.15in}
\subsection{\ac{TRN} Ranking Validation}
To evaluate the effectiveness of the proposed TRN ranking, we define two complementary performance metrics: 
\textit{Path Coverage (PC)} and \textit{Reliability Contribution (RC)}. 
These metrics jointly capture the structural and reliability-based significance of nodes in the network.

% CPC quantifies the structural importance of nodes based on their participation in the network’s shortest paths, indicating how effectively the selected TRNs enhance key distribution reachability.
% In contrast, NRC measures the reliability-weighted contribution of nodes to overall network robustness, capturing their influence on maintaining secure and resilient QKD operations.
\subsubsection{Path Coverage (PC)} PC quantifies the structural importance of nodes based on their participation in the network’s shortest paths, indicating how effectively the selected TRNs enhance key distribution reachability.

Let $P_{s,t}$ represent the shortest path between source $s$ and destination $t$.  For a given node $v$, the function $\delta_{s,t}(v)$ equals 1 if $v$ lies on $P_{s,t}$ (excluding $s$ and $t$), and 0 otherwise.  For node $v$, let
\begin{equation}
    C_v = \sum_{s,t \in V, s \neq t} \delta_{s,t}(v), \quad
    \delta_{s,t}(v) = 
    \begin{cases}
    1,& v \in P_{s,t} \setminus \{s,t\},\\
    0,& \text{otherwise}.
    \end{cases}
\end{equation}
The normalized value of the path coverage for node $v$ is
\begin{equation}
    \text{PC}_v = \frac{C_v}{\sum_{j \in V} C_j}.
\end{equation}

\subsubsection{Reliability Contribution (RC)}
RC measures the reliability-weighted contribution of nodes to overall network robustness, capturing their influence on maintaining secure and resilient QKD operations.

\begin{figure}[t]
    \centering
    \includegraphics[width=0.45\textwidth]{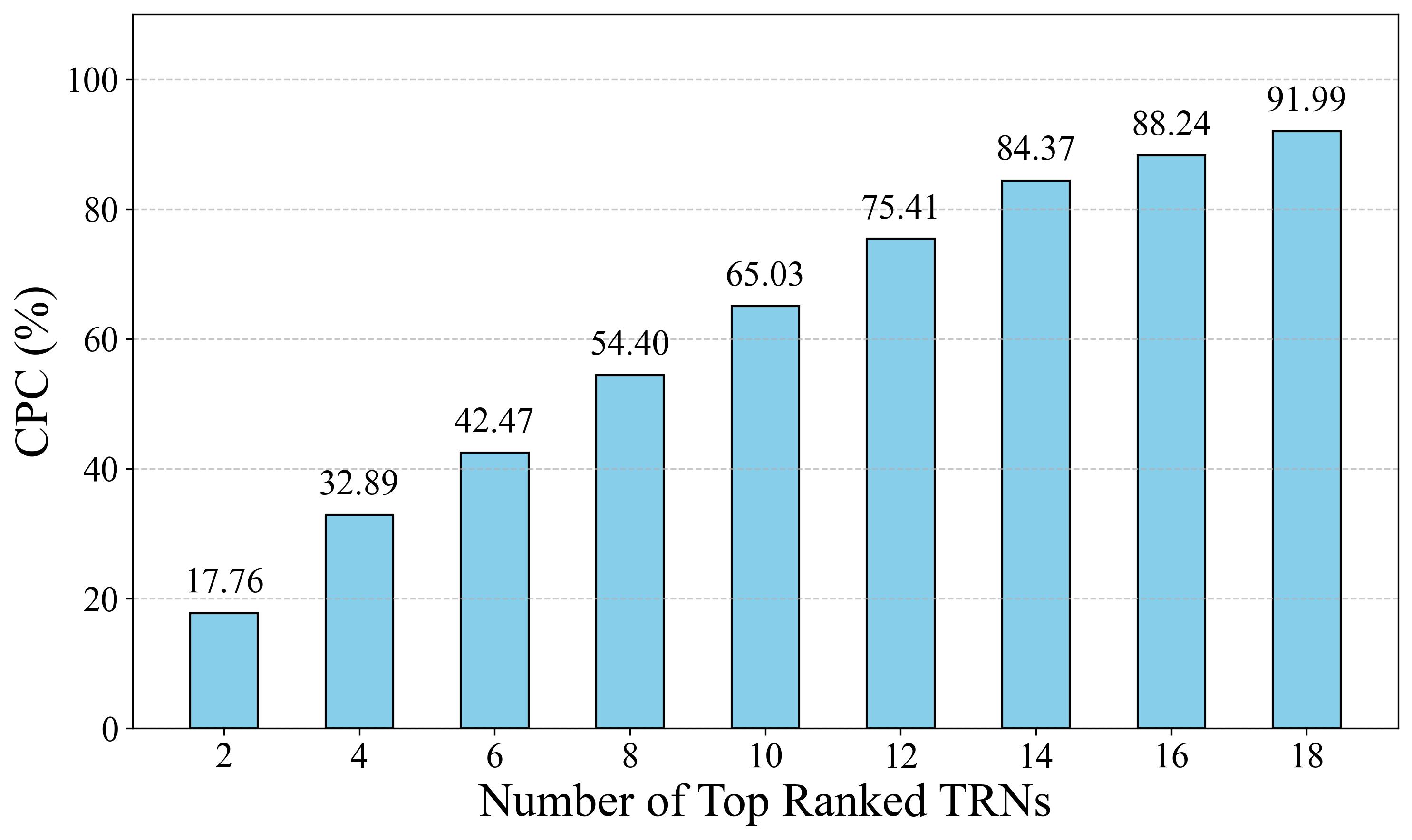}
    \vspace{-.1in}\caption{Cumulative path coverage (CPC) by \acp{TRN} according to total score.}
    \label{cpc_vs_trns_total_score}
\end{figure}

\begin{figure}[t]
    \centering
    \includegraphics[width=0.45\textwidth]{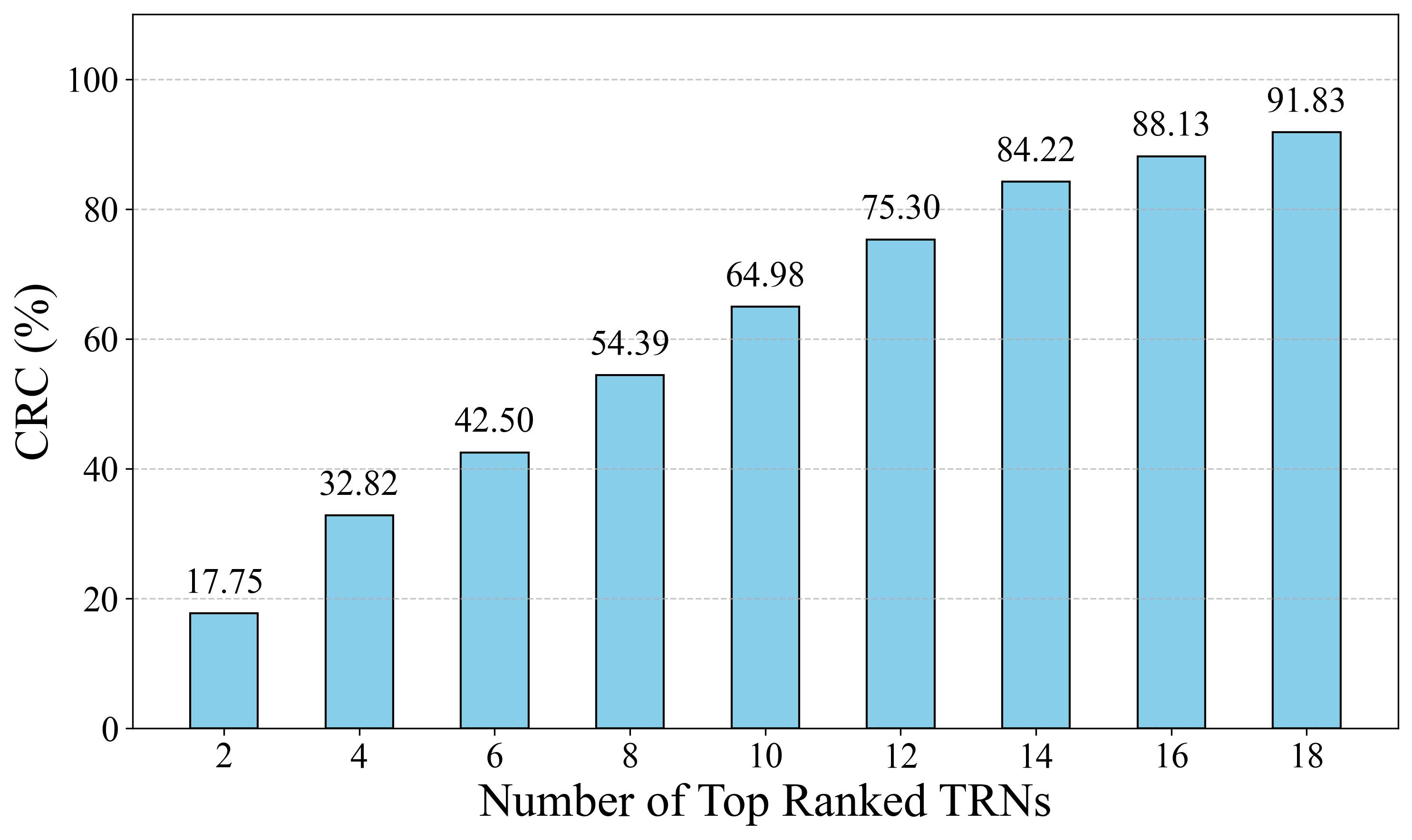}
    \vspace{-.1in} \caption{Cumulative Reliability Contribution (CRC) by \acp{TRN} according to total score.}
    \label{crc_vs_trns_total_score}
\end{figure}

The reliability of a node $v$, denoted $R_{v}$, is obtained from the Bayesian reliability framework described earlier. 
  Let $\mathcal{P}_v$ denote the set of shortest paths of all connections traversing node $v$. The reliability of a path $P_{s,t}$ is
\begin{equation}
    R_{s,t} = \prod_{v \in P_{s,t} \setminus \{s,t\}} R_{v}
\end{equation}
The overall reliability contribution of node $v$ is
\begin{equation}
    R_{\text{overall}}^{(v)} = \sum_{(s,t) \in \mathcal{P}_v} R_{s,t},
\end{equation}
and the normalized value of the reliability contribution is
\begin{equation}
    \text{RC}_v = \frac{R_{\text{overall}}^{(v)}}{\sum_{i \in V} R_{\text{overall}}^{(i)}}.
\end{equation}
Figures~\ref{cpc_vs_trns_total_score} and~\ref{crc_vs_trns_total_score} show the cumulative path coverage (CPC) and cumulative reliability contribution (CRC), respectively, achieved by the top-ranked \acp{TRN} as a function of the number of selected top-ranked \acp{TRN}. Both metrics follow a similar increasing trend, demonstrating that nodes with higher reliability contributions also occupy structurally critical positions within the network. The CPC curve rises sharply for the initial few nodes, covering approximately $75\%$ of all shortest paths with only the top 12 \acp{TRN}. 
Beyond this point, the marginal improvement decreases because the majority of shortest paths have already been covered by the highest-ranked nodes, and additional TRNs mainly introduce redundancy rather than new coverage.
Similarly, the CRC curve exhibits a comparable trend.

These results collectively confirm that a relatively small subset of highly ranked nodes, typically the top 7--8, achieves a balanced trade-off between coverage and reliability. Consequently, the combined analysis of CPC and CRC validates that the proposed reliability-aware ranking framework effectively identifies nodes that are both topologically central and operationally trustworthy, improving the reliability and robustness of TRN selection.

% Figures~\ref{cpc_vs_trns_total_score} and \ref{nrc_vs_trns_total_score} show that both CPC and CRC metrics exhibit a similar increasing trend, indicating that nodes contributing most to reliability also occupy critical topological positions. The CPC curve rises sharply for the first few nodes, covering nearly 53\% of all shortest paths with only the top eight \acp{TRN}. Beyond this point, the curve saturates, suggesting that additional \acp{TRN} contribute marginal gains in path coverage. In contrast, CRC continues to grow more gradually, showing that the inclusion of slightly lower-ranked nodes still enhances the overall network reliability.

% Together, these observations confirm that a small subset of high-ranked nodes typically the top 7–8, achieves a balanced trade-off between coverage and reliability. Hence, CRC and CPC jointly validate that the proposed reliability-aware ranking effectively identifies structurally central and operationally trustworthy \ac{TRN} candidates, ensuring secure and resilient QKD key relay.

\subsection{Comparative Analysis}
To further validate the proposed ranking model, we compare its performance against a degree centrality-based heuristic using both CPC and CRC as evaluation metrics. Degree centrality is selected as the baseline because it is one of the simplest and most widely adopted graph-theoretic heuristics for identifying structurally important nodes, making it a suitable reference for evaluating the effectiveness of the proposed reliability-aware ranking framework.

\begin{figure}[t]
    \centering
    \includegraphics[width=0.45\textwidth]{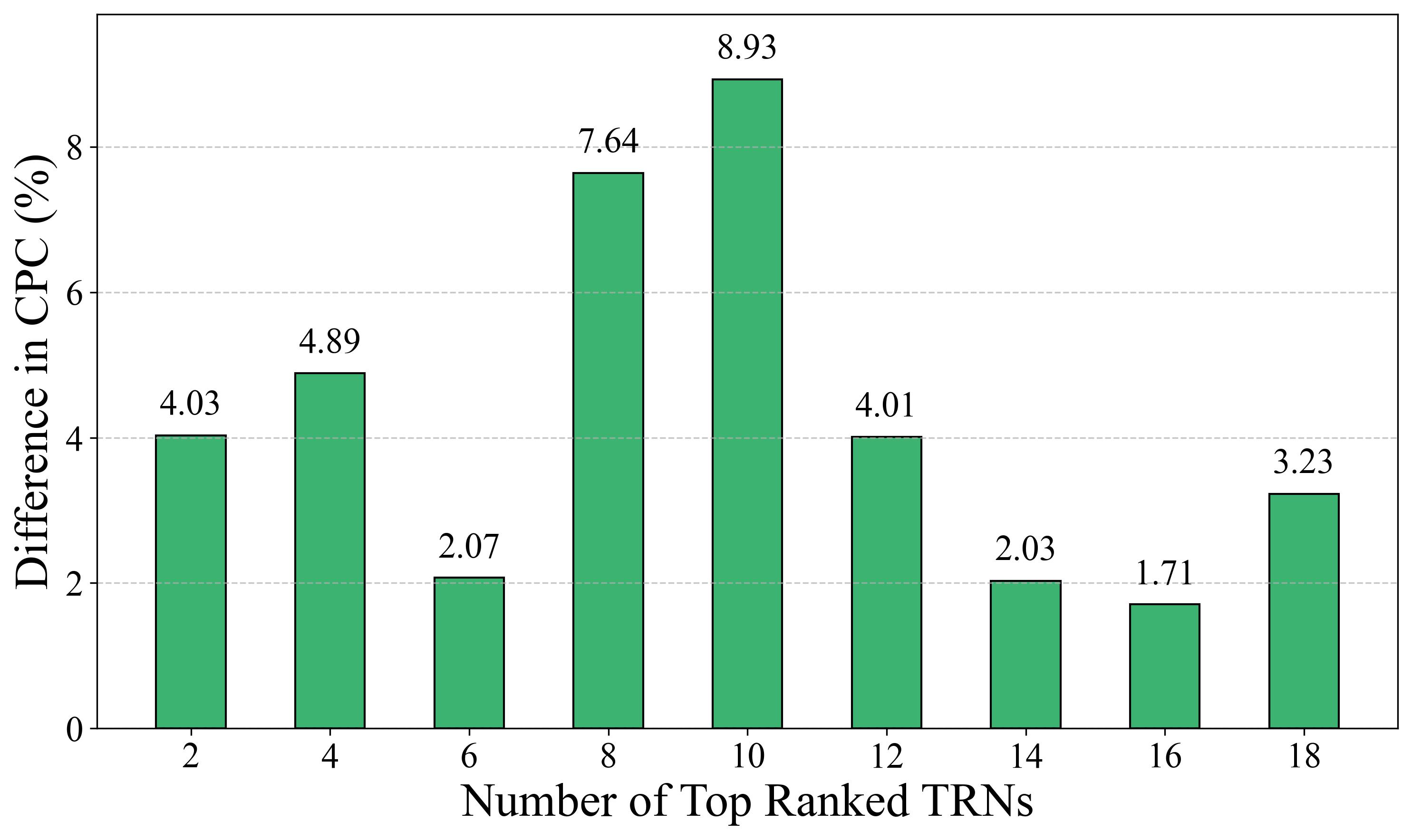}
    \caption{Difference in CPC between proposed composite score and degree centrality-based \ac{TRN} selection.}
    \label{cpc_vs_trns_diff}
\end{figure}

As shown in Fig.~\ref{cpc_vs_trns_diff}, the proposed approach consistently outperforms degree centrality across all values of $K$, with the most pronounced improvement observed for intermediate selections. For instance, at $K=10$, the difference in CPC reaches a maximum of 8.93\%, while for $K=8$ and $K=12$, the improvements are 7.64\% and 4.01\%, respectively. This demonstrates that the top-ranked nodes identified by our reliability-aware model not only occupy critical network positions but also contribute more significantly to overall reliability aggregation. A similar trend is also observed for CRC, where the proposed ranking achieves superior path coverage across nearly all $K$ values. The gain remains most prominent throughout the range, highlighting the ability of the proposed scoring function to select structurally and reliability-relevant nodes that maximize network coverage.

Overall, both metrics reveal that while degree centrality captures basic connectivity, it fails to account for node reliability and global path influence. The proposed composite score, integrating eigen vector and betweenness centralities with node reliability, results in \ac{TRN} selections that are both structurally strategic and reliability-efficient, offering superior performance for mid-range \ac{TRN} counts (approximately 7–12 \acp{TRN}).

\begin{figure}[t]
    \centering
    \includegraphics[width=0.45\textwidth]{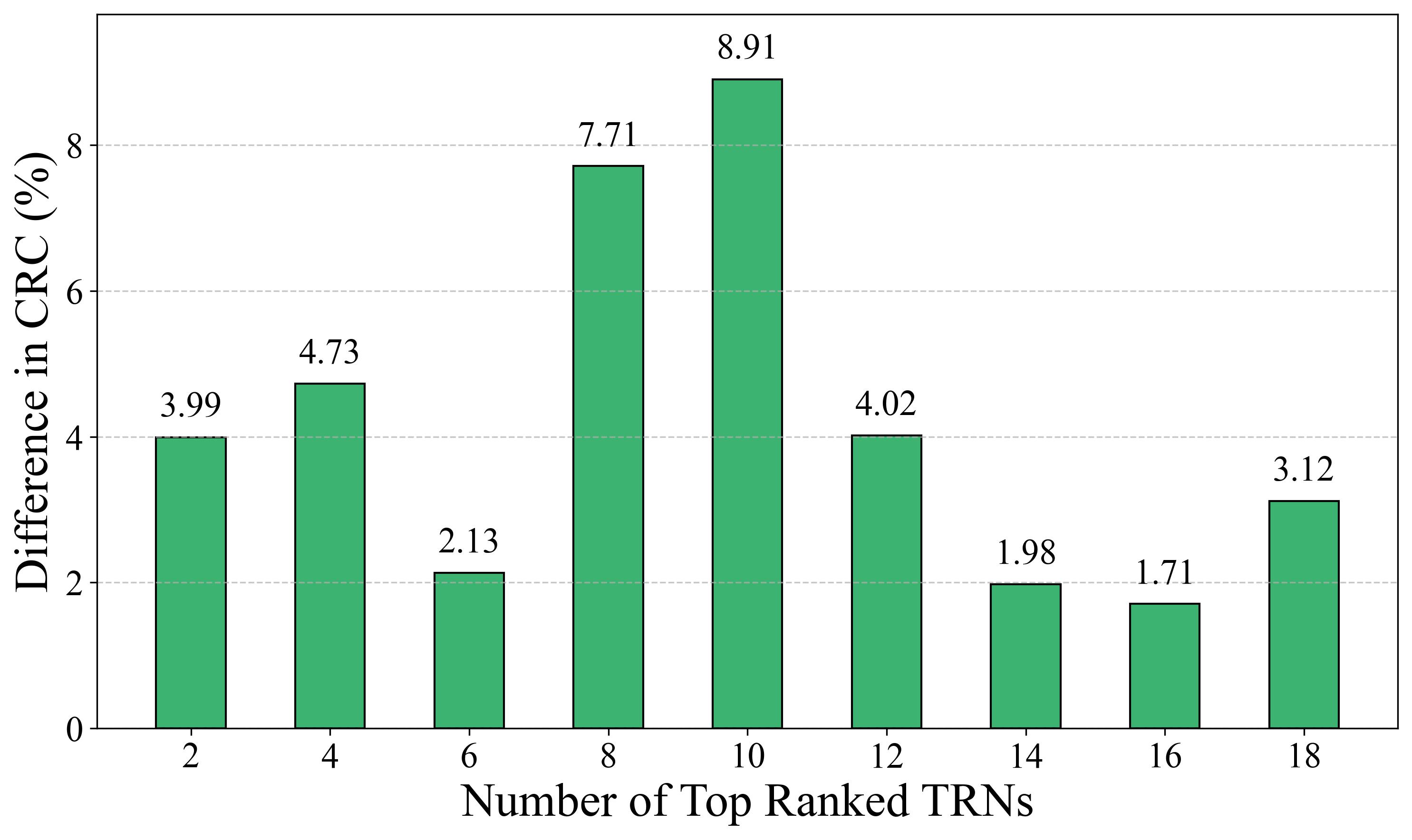}
    \caption{Difference in CRC between proposed composite score and degree centrality-based \ac{TRN} selection.}
    \label{nrc_vs_trns_diff}
\end{figure}

% The present study assumes that node reliability indicators are obtained through simulation rather than from measurements collected in operational QKD deployments. Nevertheless, the proposed Bayesian framework is generic and can directly incorporate real monitoring information, such as security logs, maintenance records, and operational statistics, whenever such data become available.

% \section{Conclusions}
% In this work, we proposed a reliability-aware \ac{TRN} selection framework to enhance secure and resilient QKD operations over metro optical networks. A Bayesian fusion-based reliability model was introduced to quantify the trustworthiness of each node by aggregating multiple operational and security indicators into a unified probabilistic reliability score. This reliability score was then incorporated into the network through reliability-weighted link cost modification in the Dijkstra algorithm, enabling the identification of shortest and most reliable paths. Subsequently, a composite total score combining eigenvector and betweenness centralities was used to rank candidate \acp{TRN}. Simulation results on a realistic 28-node metro topology demonstrate that the proposed framework achieves up to 8.91\% higher reliability contribution and 8.93\% greater path coverage compared to degree centrality. The joint use of Bayesian reliability estimation and topology-aware ranking ensures reliable, topology-aware, and resilient TRN selection for practical QKD-enabled optical networks.

\section{Conclusions}

In this work, we proposed a reliability-aware TRN selection framework for QKD-enabled metro optical networks. A Bayesian fusion-based reliability model was combined with reliability-aware shortest-path computation and graph centrality measures to rank candidate TRNs. Simulation results on a 28-node metro topology demonstrate up to 8.91\% higher reliability contribution and 8.93\% greater path coverage than degree centrality, highlighting the effectiveness of the proposed framework. Future work will investigate dynamic TRN selection under varying traffic demands.

%Future work involves formulating an optimization framework that jointly considers traffic demands, node reliability, and key generation constraints for dynamic TRN selection.
\begin{acronym}
    \acro{QKD}{Quantum Key Distribution}
    \acro{TRN}{Trusted Repeater Node}
\end{acronym}

\bibliographystyle{IEEEtran}
\bibliography{ref}
% \bibliography{IEEEabrv, ref.bib}
\end{document}